\documentclass[aps,pra,preprint,groupedaddress]{revtex4}
\usepackage{graphicx}

\begin{document}

\title{The evolution of additional (hidden) quantum variables in the interference
of Bose-Einstein condensates}
\author{W. J. Mullin$^{a}$, R. Krotkov$^{a}$, and F. Lalo\"{e}$^{b}$}
\affiliation{$^{a}$Department of Physics, University of Massachusetts,
Amherst, Massachusetts 01003 USA\\
$^{b}$LKB, Dept.\ de Physique de l'ENS, 24 rue Lhomond, 75005, Paris, France}

\date{\today}

\begin{abstract}
Additional variables (also often called ``hidden variables'') are sometimes
added to standard quantum mechanics in order to remove its indeterminism or
``incompletness,'' and to make the measurement process look more classical.
Here we discuss a case in which an additional variable arises almost
spontaneously from the quantum formalism: the emergence of relative phase
between two highly populated Fock state Bose-Einstein condensates. The model
simulated here involves the interference of two Bose condensates, one with
all up spins, and the other with down spins, along a z-axis. With the clouds
overlapping, we consider the results of measuring spins in a transverse
plane (the general direction is studied in an appendix). The determination
of the previously ``hidden'' phase becomes progressively more definite as
additional measurements are made. We also provide an analysis of a recent
and closely related experiment.
\end{abstract}

\maketitle

\section{Introduction}

Several laboratories have made use of Bose condensates with spins or
pseudo-spins to perform very interesting experiments. For example, in JILA
experiments\cite{JILA},\cite{JILA2}, a mixture of two hyperfine states
playing the role of pseudo-spin-1/2 particles was found to segregate by
species, thereby exhibiting spin waves. Another ``spin'' experiment\cite
{Hall} involves overlapping two Bose condensates, one with spin up and the
other with spin down and observing the appearance of a spontaneous
transverse spin polarization. In this paper we consider a simulation that
involves the interference of two clouds of Bose gases, one with up spins and
the other down spins along some z-axis. With the clouds overlapping, we
consider what happens in measuring spins along a transverse direction at a
set of azimuthal angles, detecting whether a spin is found up or down along
each measurement angle used. (The case of a general measurement direction is
studied in Appendix A.) In looking at a simulation of this process we note
an interesting element involved in the process: the relation of the
successive measurements to the appearance of a so-called ``hidden'' variable.

The introduction of additional variables to the standard formalism of
quantum mechanics is not a new idea; it dates back almost to the appearance
of this theory \cite{Jammer}, with, for instance, the early work of L. de
Broglie\cite{De Broglie} and later D.\ Bohm on ``hidden variables''\cite
{Bohm}.\ The main motivation was to restore determinism by reproducing the
statistical results of quantum mechanics with classical averages over
additional variables. In 1935, Einstein, Podolsky and Rosen (EPR) showed
that there exists an even stronger argument to complete quantum mechanics,
without referring to determinism: the standard form of quantum mechanics
does not satisfy local realism, and should be completed by additional
``elements of reality''\ to restore it \cite{EPR}.\ Thirty years later J.\
Bell, with a famous theorem \cite{Bell1, Bell2}, extended the EPR\ argument
and showed that, not only the formalism, but also the predictions of quantum
mechanics are sometimes incompatible with local realism.\ In other words,
adding variables to quantum mechanics is not sufficient to restore local
realism in all cases: there exist some situations where the evolution of
additional variables has to be explicitly non-local. This remarkable result
stimulated several generations of careful experiments in order to decide
whether or not quantum mechanics still gives correct predictions, even in
these surprising ``non-local''\ situations.\ An impressive body of evidence
has now been accumulated in favor of the predictions of quantum mechanics,
even if none of these experiments is ideal\cite{footnote}
; the general consensus among physicists is that these surprising
``non-local''\ predictions of quantum mechanics are indeed obeyed by
Nature.\ This does not mean that additional variables should now be excluded
from quantum mechanics! Actually, advocates of these variables argue that it
is precisely one of their big merits to make the non-local character of
quantum mechanics explicit \cite{Goldstein}. For a general discussion of the
consequences of the Bell theorem, see for instance Refs.~\onlinecite{Mermin,
AJP}.

In most cases, the introduction of additional variables into quantum
mechanics leaves a large range of flexibility: one has to introduce whatever
new variables seem appropriate (positions, momenta, fields, etc.) with
adequate equation of evolution, chosen so that a statistical average over
initial condition restores the usual predictions of quantum mechanics. There
is nevertheless a case where an additional variable emerges almost
spontaneously, in an unique way, from the quantum formalism: the spontaneous
appearance of relative phase between two highly populated Fock states
(Bose-Einstein condensates).\ The general physical phenomenon was discussed
long ago by Anderson \cite{Anderson}, with more emphasis on spontaneous
symmetry breaking in phase transitions than additional variables in quantum
mechanics.\ Later several authors used various formalisms, often borrowed
from quantum optics, to give more detailed calculations\cite{Javanainen, M2}%
.\ The approach used in Ref.~\onlinecite{Phase} is slightly different since
it shows how a simple conservation rule, the conservation of particle
number, can naturally be expressed through an integral over the
``conjugate''\ variable, the relative phase $\Phi ~$of the two states.\ The
probability of any sequence of results then appears as a sum over this new
variable, exactly as in theories with additional variables.\ Moreover, each
time a measurement is performed, the state vector projection postulate
provides a new initial state, which changes the $\Phi $ distribution; it
turns out that the change of this probability distribution can be obtained
very easily for any sequence of measurements.\ In other words, one has
access to the evolution of the distribution function of the additional
variable under the effect of one or more successive quantum measurements.
This is what we study in the present article; for the sake of simplicity, we
limit ourselves to the limit of large occupation numbers, a case in which
the Bell theorem does not predict any incompatibility between quantum
mechanics and local realism, so that locality will not be an issue here.

Additional variables in quantum mechanics are often called ``hidden
variables''\ for historical reasons.\ \ Neverheless, J.\ Bell pointed out
how inappropriate this name is (``Absurdly, these theories are known as
hidden variable theories....''\ )\cite{Bell3}, pointing out that the name
would be more appropriate for the standard wave function of quantum
mechanics.\ Indeed, in an interference experiment, for instance, these
additional variables are not hidden but actually directly observed in the
result of the individual experiments.\ On the other hand, the wave function
or state vector can be reconstructed only indirectly by statistical analysis
after many measurements.\ This is why we will tend to avoid the words
``hidden variables''\ and rather speak of ``additional variables''\ here.

\section{Phase in spin states}

Suppose we have particles with two internal states, either real spin 1/2 or
pseudo-spins as in the case of two hyperfine states. If we have two
stationary clouds of Bose particles with $N_{+}$ spin-up, and $N_{-}$
spin-down, particles along a $z$-axis, the initial state is 
\begin{equation}
\left| \Psi \right\rangle =\left| N_{+},N_{-}\right\rangle ,
\label{InitState}
\end{equation}
a Fock state in spin space. We want to measure the occurrences of a sequence
of spin measurements in the transverse $xy$-plane, at a series of azimuthal
angles, $\phi _{1},$ $\phi _{2},$ $\cdots \phi _{m},$ resulting in the
sequence of results $\left\{ \eta _{i}\right\} ,$ either up (+) or down ($-)$
along the angles: 
\begin{equation}
\eta _{1}=\pm 1;\eta _{2}=\pm 1;\cdots \eta _{m}=\pm 1.
\end{equation}
Here we present a simplified calculation of the probabilities, which
completely ignores orbital variables; a more precise calculation is given in
Appendix A and Ref.~\onlinecite{Phase}. We have a set of angular momentum
variables given in terms of the destruction operators $a$ for particles up,
and $b$ for down, along a $z$-axis. The number operator is then 
\begin{equation}
n=a^{\dagger }a+b^{\dagger }b,
\end{equation}
with the angular momentum variables 
\begin{eqnarray}
\sigma _{z} &=&a^{\dagger }a-b^{\dagger }b,  \nonumber \\
\sigma _{x} &=&a^{\dagger }b+b^{\dagger }a,  \nonumber \\
\sigma _{y} &=&i(b^{\dagger }a-a^{\dagger }b).  \label{spinops}
\end{eqnarray}

These definitions should make clear the meaning of the axes of our
$xyz$ coordinate system and the term "transverse."  The z-axis is
defined by the two states (possibly hyperfine states) into which our
particles are condensed in Eq.(1).  Thus $\sigma_{z}$ is diagonal in
Fock states, while $\sigma_{x}$ and $\sigma_{y}$ are not.  For
example, while $\left| 1,0 \right\rangle$ as an eigenstate of
$\sigma_{z}$, an eigenstate of $\sigma_{x}$ is the mixed state $\left|
1,0\right\rangle +\left| 0,1 \right\rangle$.

The expectation value of the operator 
\begin{eqnarray}
p_{\eta }(\phi _{i}) &=&\frac{1}{2N}\left[ n+\eta \left( \cos \phi
_{i}\sigma _{x}+\sin \phi _{i}\sigma _{y}\right) \right]  \nonumber \\
&=&\frac{1}{2N}\left[ n+\eta \left( e^{i\phi_{i} }b^{\dagger }a+e^{-i\phi_{i}
}a^{\dagger }b\right) \right]
\end{eqnarray}
gives the probability of finding a spin with component $\eta $ along the
angle $\phi _{i}$ in the transverse plane. (To see the connection between
this operator and the probability, see Appendix A or Ref.~\onlinecite{Phase}%
.) Suppose one measures a single spin along a transverse axis at angle $\phi
_{1}$ starting in the state $\left| \Psi \right\rangle $ quoted above. The
probability of finding spin with result $\eta _{1}$ is easily seen to be 
\begin{equation}
P_{1}(\phi _{1})=\left\langle N_{+},N_{-}\right| \frac{1}{2N}\left[ n+\eta
_{1}\left( e^{i\phi _{1}}b^{\dagger }a+e^{-i\phi _{1}}a^{\dagger }b\right)
\right] \left| N_{+},N_{-}\right\rangle =\frac{1}{2}
\end{equation}
as expected. However, if immediately after, one measures a second particle
along a different transverse axis $\phi _{2}$ then a straightforward
calculation gives the probability of the sequence \{$\eta _{1},\eta _{2}\}$
to be 
\begin{eqnarray}
P_{2}(\phi _{1},\phi _{2}) &=&\left\langle N_{+},N_{-}\right| p_{\eta
_{2}}(\phi _{2})p_{\eta _{1}}(\phi _{1})\left| N_{+},N_{-}\right\rangle 
\nonumber \\
&=&\frac{1}{4}\left[ 1+\eta _{1}\eta _{2}\frac{x^{2}}{2}\cos (\phi _{1}-\phi
_{2})\right] ,
\end{eqnarray}
where $x=2\sqrt{N_{+}N_{-}}/N$ and $N=N_{+}+N_{-}.$ Having found the first
particle up or down along $\phi _{1}$ affects the result of the second
measurement along $\phi _{2}.$ The two measurements are correlated by boson
statistics. It is possible to write this last result in an instructive way
as 
\begin{equation}
P_{2}(\phi _{1},\phi _{2})=\frac{1}{4}\int_{0}^{2\pi }\frac{d\Phi }{2\pi }%
\left( 1+x\eta _{1}\cos (\phi _{1}-\Phi )\right) \left( 1+x\eta _{2}\cos
(\phi _{2}-\Phi )\right) ,
\end{equation}
as one can verify by doing the integration. It is remarkable that this form
of the probability for a sequence of $m$ such measurements along transverse
axes $\phi _{1},$ $\phi _{2},$ $\cdots \phi _{m}$ persists\cite{Phase}:

\begin{equation}
P_{m}\sim \int_{0}^{2\pi }\frac{d\Phi }{2\pi }\prod_{i=1}^{m}\left[ 1+x\eta
_{i}\cos (\phi _{i}-\Phi )\right] .  \label{Main}
\end{equation}

This result just quoted assumes all measurements are done in the transverse
plane. However, measurements could also be done of spin up and down along an
arbitrary axis at angles ($\theta _{i},\phi _{i}).$ The generalization of
Ref.~\onlinecite{Phase} to this case is described in Appendix A. The
derivation there, when specified to all $\theta _{i}=\pi /2$ and uniform
orbital variables, reduces to Eq.~(\ref{Main}) as expected.

The form of Eq.~(\ref{Main}) is quite interesting. Suppose we had started
out with a system polarized partially (at a fraction $x)$ along the
transverse direction $\Phi ;$ then the resulting probability for finding a
set of spins up or down along the set of angles $\{\phi _{i}\}$ would be Eq.
(\ref{Main}) but \emph{without }the integration over $\Phi .$ Since we
started out knowing the numbers of particles in longitudinal states, all
transverse phases must be present equally in the initial state as
represented by the integral over the transverse phase in Eq.~(\ref{Main}).

We can look at the expression of Eq.~(\ref{Main}) is a somewhat different
way: The probability that the $\eta $ in the $m$th spin measurement is $\pm
1,$ after a sequence of results $\{\eta _{i}\},$ is 
\begin{equation}
P_{m}(\pm )\sim \int_{0}^{2\pi }d\Phi \;g_{m}(\Phi )\left( 1\pm x\cos (\phi
_{m}-\Phi )\right)  \label{SpinIntegral}
\end{equation}
where the $m$th measurement is made only in the transverse plane at the
angle $\phi _{m}.$ In this we have 
\begin{equation}
g_{m}(\Phi )=\prod_{i=1}^{m-1}\left( 1+x\eta _{i}\cos (\phi _{i}-\Phi
)\right) .
\end{equation}
The function $g_{m}(\Phi )$, as we will see by explicit simulation, peaks
up, after a sufficiently large number $m-1$ of measurements, at some value
of phase, call it $\Phi _{0}.$ Thus subsequent measurements will appear as
if the the system of particles had, after just $m-1$ measurements (with $%
m\ll N$), been prepared with a set polarization phase angle. (Of course, if
we repeated the experiment, starting again from the same initial state, a
different phase would emerge and indeed it is random in a series of such
experiments.) Because particle number and phase are conjugate variables, the
original state of known particle number has, after $m$ measurements, morphed
into a state of relatively well-known phase, but with the number of
particles up or down along $z$ much less certain (although the total number
or particles remains known).

We start with $\left| \Psi \right\rangle =\left| N_{+},N_{-}\right\rangle .$
Our first measurement along some azimuthal angle $\phi $ produces the new
state 
\begin{equation}
\left| \Psi _{1}\right\rangle =\frac{1}{2}(n+\eta \left( e^{i\phi
}b^{\dagger }a+e^{-i\phi }a^{\dagger }b\right) )\left|
N_{+},N_{-}\right\rangle .
\end{equation}
It is clear that this process produces a mixture of states $\left|
N_{+},N_{-}\right\rangle ,$ $\left| N_{+}+1,N_{-}-1\right\rangle ,$ and $%
\left| N_{+}-1,N_{-}+1\right\rangle ,$ Each subsequent measurement produces
a further mixing of number states so that number becomes less certain. Its
conjugate, phase, becomes less uncertain.

As an alternative view one might think of the phase as having been there all
along, but temporarily hidden from view, with experiments continually
clarifying its value. In such a view one integrates probabilities over all
possible values of an additional variable introduced to ``complete'' quantum
mechanics, as we find in Eqs. (\ref{Main}) or (\ref{SpinIntegral}). The
additional quantum variable is this ``emerging'' phase angle.

In the rest of this section we confine our discussions to the simplified
case of $N_{+}=N_{-}=N/2$, or $x=1$. Let us now Fourier transform $%
g_{m}(\Phi )$ to write 
\begin{equation}
g_{m}(\Phi )=a_{0}^{m}+\sum_{q=1}^{\infty }\left[ a_{q}^{m}\cos (q\Phi
)+b_{q}^{m}\sin (q\Phi )\right] .  \label{origg}
\end{equation}
In Eq.~(\ref{SpinIntegral}), only the $q=1$ term contributes, so that 
\begin{equation}
P_{m}(\pm )=\frac{1}{2}\left[ 1\pm \left( \frac{a_{1}^{m}}{2a_{0}^{m}}\cos
\phi _{m}+\frac{b_{1}^{m}}{2a_{0}^{m}}\sin \phi _{m}\right) \right] .
\label{a1b1Eq}
\end{equation}

\begin{figure}[tbp]
\includegraphics[width=3.98in, height=3.55in]{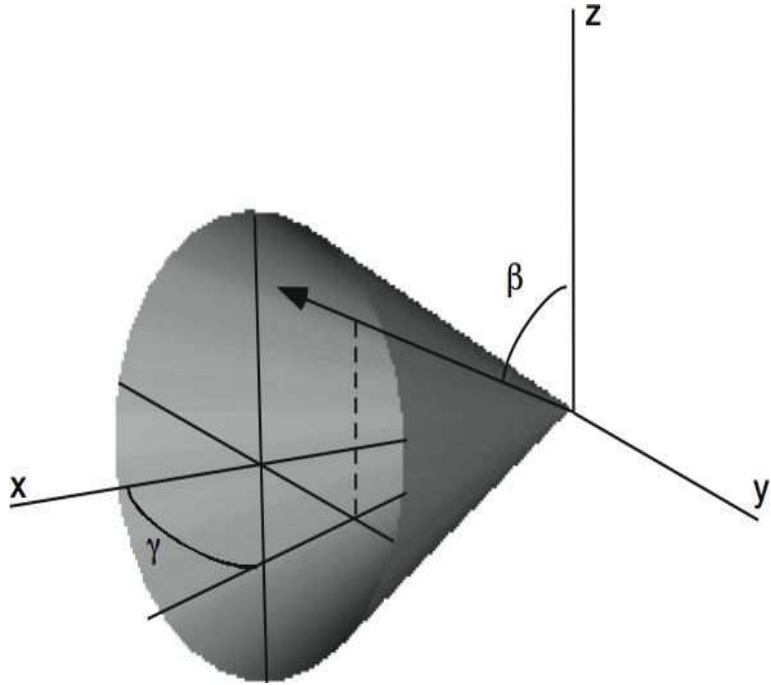}
\caption{The cone of equal probability. Any spin at angles $\beta ,\gamma $
on this cone will give the same probability of being up along the
measurement axis, taken as $x$ in this figure.}
\label{FirstCone}
\end{figure}

With this result we can define two parameters that characterize the the
results of the $m$th measurement. These parameters are $\Phi _{m}$ and $%
\alpha _{m}$ (or $A_{m}$) given by 
\begin{equation}
\cos \Phi _{m}=\frac{a_{1}^{m}}{\sqrt{(a_{1}^{m})^{2}+(b_{1}^{m})^{2}}}%
~;~\sin \Phi _{m}=\frac{b_{1}^{m}}{\sqrt{(a_{1}^{m})^{2}+(b_{1}^{m})^{2}}}%
;~~\sin \alpha _{m}=A_{m}=\frac{\sqrt{(a_{1}^{m})^{2}+(b_{1}^{m})^{2}}}{%
2a_{0}^{m}}.  \label{eqn3}
\end{equation}
(Because $P_{m}$ is a probability, $A_{m}\leq 1$ and we can write $A_{m}$ as
a sine.) We obtain 
\begin{equation}
P_{m}(\pm )=\frac{1}{2}\left[ 1\pm \sin \alpha _{m}(\cos \phi _{m}\cos \Phi
_{m}+\sin \phi _{m}\sin \Phi _{m})\right] .  \label{ProbSpin}
\end{equation}
Consider a single spin polarized as some space angle $(\beta ,\gamma )$ and
assume that a measurement if performed along a transverse direction $\phi
_{m}$.\ The probabity of finding a $+$ result is given by the well-known
expression 
\begin{equation}
\left| _{\pm }\left\langle \phi _{m}\mid \beta ,\gamma \right\rangle \right|
^{2}=\frac{1}{2}\left[ 1\pm \sin \beta (\cos \phi _{m}\cos \gamma +\sin \phi
_{m}\sin \gamma )\right] ,  \label{genProb}
\end{equation}
which has exactly the same form as Eq.~(\ref{ProbSpin}) if $\beta =\alpha
_{m}$, $\gamma =\Phi _{m}$. Actually the two expressions are equal whenever
the ``spin'' lies on a cone of directions around the measurement direction $%
\phi _{m}$ containing this particular direction. We show this ``cone of
equal probability'' (CEP) in Fig.~\ref{FirstCone} with a spin at arbitrary $%
\beta ,\gamma $ on the cone.

What this calculation shows is that two different points of view are
possible.\ At each measurement step, all the effects of the previous
measurements are contained in the distribution function $g_{m}(\Phi )$; in
this first point of view, the system is described by a statistical mixtures
of spins polarized in transverse directions.\ But one obtains the same
probabilities for the next measurement by replacing this statistical mixture
by a single pure spin state, fully polarized along directions $\beta =\alpha
_{m}$, $\gamma =\Phi _{m}$ (or any direction making the same angle with the
direction of measurement).\ A second point of view is therefore possible,
where the spin is fully polarized, but in a direction that is, in general,
no longer in the transverse plane.\ For instance, if the distribution $%
g_{m}(\Phi )$ has no $\Phi $ dependence, the $\Phi _{m}$ dependence of the
probabilities disappears, and this spin is polarized in a direction that is
perpendicular to the direction of measurement ($\alpha _{m}=0$).\ On the
other hand, the $\Phi _{m}$ dependence is maximum when $\alpha _{m}=\pi /2$
so that the pure spin state lies in the transverse plane.

\section{The Amherst Experiment}

Ref.~\onlinecite{Hall} reports an experiment that is well-described by our
analysis. In Ref.~\onlinecite{Hall} the experiment is treated in terms of a
two-component spinor, with each component representing one of the two
condensates in a hyperfine state with a well-defined phase. Here we consider
the experiment in terms of a Fock state having an up-spin-down-spin ratio
the same as the experiment. The Amherst experiment combines two condensates
each originating from a different point and claims that the result shows a
``spontaneous transverse polarization.'' However, what they are able to
detect by laser absorption is the existence of each of the longitudinal
components. So after mixing the two components they perform a $\pi /2$ tip
in order to measure the size of the polarization and its phase. What they
see then is an anticorrelated interference pattern in the two components.

We assume a Fock state for the initial mixed-condensate function: $\left|
\Psi \right\rangle =\left| N_{+},N_{-}\right\rangle ,$ with an arbitrary
pre-established $x,y,z$ coordinate system. We introduce a set of transverse
coordinates $\hat{u},\hat{w}$ offset by an angle $\phi $ relative to the
original $xy$ axes: 
\begin{eqnarray}
\hat{u} &=&\cos \phi \hat{x}+\sin \phi \hat{y} \\
\hat{w} &=&-\sin \phi \hat{x}+\cos \phi \hat{y}
\end{eqnarray}
We want to measure the $\hat{u}$-component of the spin in multiple
simultaneous measurements made along the longitudinal ($z)$ direction. To
enable this we do a spin tip of $\pi /2$ around the $\hat{w}$ axis. Then one
measures spins along $\hat{z}.$ So the wave function analyzed is 
\begin{equation}
\left| \Psi ^{\prime }\right\rangle =U\left| \Psi \right\rangle =U\left|
N_{+},N_{-}\right\rangle
\end{equation}
where 
\begin{equation}
U=e^{i\pi \sigma _{w}/4}
\end{equation}
and 
\begin{equation}
\sigma _{w}=-\sin \phi \sigma _{x}+\cos \phi \sigma _{y}
\end{equation}

After applying $U$ we make measurements along $\hat{z},$ that is, we look
for the probability given by 
\begin{equation}
P_{m}=\left( \Psi \right| U^{\dagger }\prod_{i=1}^{m}\frac{1}{2N}\left(
n+\eta _{i}\sigma _{z}\right) U\left| \Psi \right\rangle
\end{equation}
This is precisely equivalent to the relation 
\begin{equation}
P_{m}=\left( \Psi \right| \prod_{i=1}^{m}\frac{1}{2N}\left( n+\eta
_{i}U^{\dagger }\sigma _{z}U\right) \left| \Psi \right\rangle
\end{equation}

But we have 
\begin{equation}
U^{\dagger }\sigma _{z}U=\sin \phi \sigma _{y}+\cos \phi \sigma _{x}
\end{equation}
so that 
\begin{eqnarray}
P_{m} &=&\left\langle N_{+},N_{-}\right| \prod_{i=1}^{m}\frac{1}{2N}\left(
n+\eta _{i}\left( \sin \phi \sigma _{y}+\cos \phi \sigma _{x}\right) \right)
\left| N_{+},N_{-}\right\rangle \\
&=&\frac{1}{2^{m}}\int_{0}^{2\pi }d\Phi \;g_{m}(\Phi )(1+\eta _{m}x\cos
(\phi -\Phi ))
\end{eqnarray}
where the last line follows from the discussion of the previous section. We
have seen that if we make simultaneous measurements over the whole cloud of
particles $g_{m}$ ultimately peaks sharply at a particular phase $\Phi _{0}$
, so that 
\begin{equation}
P_{m}(\eta _{m})\approx \frac{1}{2}(1+\eta _{m}x\cos (\phi -\Phi _{0}))
\label{FinalAmh}
\end{equation}
which is just the result given in Eq.~(6) of Ref.~\onlinecite{Hall}. What
this shows is that the up and down measurements will be anticorrelated in
phase. Ref.~\onlinecite{Hall} uses this anticorrelation in phase in the 
\emph{spatial} interference patterns of the two detected clouds. That
reference assumed two point sources of the two bosonic clouds released from
traps. The discussion of Appendix A does indeed include a spatial phase; if
we use $\theta _{i}=\pi /2$ there, we find we find that a spatial phase $\xi
(r)$ should be added to the argument of the cosine in Eq.~(\ref{FinalAmh}),
where $\xi (r)=\arg\left[ u_{a}(\mathbf{r})/u_{b}(\mathbf{r})\right] $ and $%
u_{a}(\mathbf{r})$ and $u_{b}(\mathbf{r})$ are the single particle wave
functions describing the two condensates. An order of magnitude of the
behavior of the interference fringes can be derived by assuming we are
allowed to substitute simple one-dimensional spreading Gaussian wave packets
for $u_{a}(\mathbf{r})$ and $u_{b}(\mathbf{r})$. (Appendix A does not
explicitly mention time dependence in the wave functions but that inclusion
is straightforward.) When we do this we find that the time-dependent phase
is of order $\xi (r,t)=md^{2}/\hbar t,$ where $d$ is the separation of the
centers of the two wave packets, $m$ the particle mass, and $t$ is the time
after the initially narrow packets have been spreading. This feature of
time-dependent fringes is assumed in the Amherst experment as it was in Ref.~%
\onlinecite {Ketterle}. A complete derivation of this fringe result is given
by Wallis et al.\cite{Wallis}

Our interpretation of the Amherst experiment shows the ambiguity of
interpretation that can arise in such an experiment. The interpretation
given by Ref.~\onlinecite{Hall} is that the spontaneous polarization existed
in the transverse plane before the ninety degree tip took place. But did it?
Before the longitudinal measurements, did $\Phi _{0}$ exist? One might
consider it an additional variable that was already there, but needed to be
brought out by experiment. Or one can say it did not exist; the wave
function was a linear combination of all such phases and was collapsed by
longitudinal measurement to the final value $\Phi _{0}$ found in the end.
Either interpretation works.

\section{Numerical Simulation}

The physics becomes clearer if we do a numerical simulation of the
measurements. The first spin result is chosen randomly up or down along $%
\phi _{1}.$ To choose from the probability $P_{m}$ at each subsequent stage (%
$m=2,3,\cdots )$ we simply pick a random number $p$ from 0 to 1. If $p<$ $%
P_{m}(+),$ then $\eta _{m}=+1$; if $p>P_{m}(+),$ then $\eta _{m}=-1.$

The first set of experiments is performed with all measurements done at the
same angle $\phi _{i}=0.$ However, we find that to gain any more information
about the emerging phase angle we must switch measurement angles.

\subsection{ Experiment with constant direction of measurement.}

Our results are shown in the first figures for a run of 300 trials with all $%
\phi _{i}\equiv \phi _{1}=0$. There is no plot of the azimuthal angle $\Phi
_{m}$ because generally in the case of a constant measurement angle, $\Phi
_{m}$ will be equal to $\phi _{1}$ or $\phi _{1}+\pi $ for \emph{every}
iteration. This is easy to understand physically by a symmetry argument:
with a constant direction of measurement, there is no way to distinguish
between spins polarized in two directions that are symmetricial with respect
to the direction of measurement in the transverse plane. With $\phi _{1}=0,$
(or consider the symmetry in terms of $\Phi ^{\prime }=\Phi -\phi _{1}),$ $%
g_{m}(\Phi )$ must be a symmetric function around $\Phi =0;$ by Eqs. (\ref
{origg}) and (\ref{eqn3}) we see $b_{1}^{m}=0,$ $\sin \Phi _{m}=0,$ and so $%
\Phi _{m}=0$ or $\pi .$ For an alternative view of the same result, look at
Eq.~(\ref{ProbSpin}) in the case of all $\phi _{m}$ identically equal to $%
\phi _{1}$. This equation is of the form 
\begin{equation}
P_{m}(\pm )=\frac{1}{2}\left[ 1\pm \sin \alpha _{m}\cos (\phi _{1}-\Phi
_{m})\right]  \label{FinalForm}
\end{equation}
But every cosine in Eq.~(\ref{SpinIntegral}) is a function of $\Phi -\phi
_{1},$ and so the integral is independent of $\phi _{1},$ as one finds by
changing integration variable to $\Phi -\phi _{1}.$ From Eq.~(\ref{FinalForm}%
) this implies that $\Phi _{m}-\phi _{1}=\mathrm{constant,}$ which we show
in Appendix B must be $0$ or $\pi $.

\begin{figure}[h]
\centering
\includegraphics[width=4in, height=2.51in]{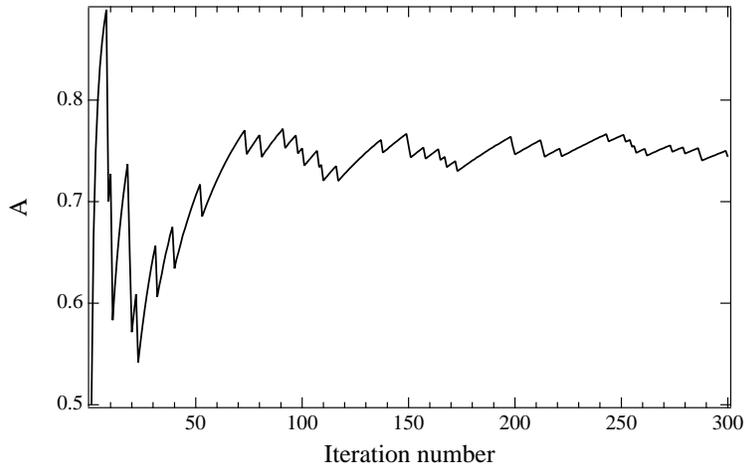}
\caption{The amplitude $A=\sin \alpha $ as a function of interation step
when all measurements have been made at a single angle. The final asymptotic
value is random.}
\label{AmpSpin}
\end{figure}

Let $m_{\pm }$ be the number of spins found with $\eta =\pm 1$ along $\phi
_{1}.$ We must always have just $P_{m}(+)=\frac{1}{2}(1\pm \sin \alpha
_{m}). $ Since we anticipate physically that $P_{m}(+)\cong
m_{+}/(m_{+}+m_{-})$ for large $m,$ the plus sign occurs when $m_{+}>m_{-}$
and the minus sign in the opposite case. In this constant $\phi _{1}$ case,
the variable $A_{m}=\sin \alpha _{m} $ goes to some unpredictable random
positive value as seen in Fig.~\ref{AmpSpin}. (As it does so, it has
slightly curved trajectories followed by small jumps: the trajectories occur
during a series of identical $\eta $ values; a jump happens when the
opposite $\eta $ eventually occurs. We explain this peculiar behavior in
Appendix C.) Fig.~\ref{ggSpin} shows $g_{300}(\Phi )$ versus $\Phi $ (as
well as the earlier and wider $g_{10}$ and $g_{150}$). It peaks up sharply
at \emph{two} locations, near $\pm \Phi _{0}$ with $\Phi _{0}=0.73$,
symmetrically situated around azimuthal measurement angle $\phi _{1}$ $=0$;
this is exactly what we should expect, as mentioned above, since, in keeping
the same measurement angle, the probability is identical for spins
symmetrically located to the sides of measurement angle. 

There is another piece of data from the run, the number of $\eta $'s of each
sign. We find that there were $m_{+}=262$ up spins and $m_{-}=38$ down spins
in the experiment. When $m_{+}>m_{-}$ as we have here, we have approximately 
\begin{equation}
\frac{1}{2}(1+\sin \alpha _{m})\cong \frac{m_{+}}{m}  \label{sinform}
\end{equation}
with $m=m_{+}+m_{-}.$ Solving we get $\sin \alpha =0.75,$ which is very
close to what we get from the limiting value in Fig. \ref{AmpSpin}. In
Appendix C we show that one can do the $g$ integral analytically for
constant measurement angle, and the actual results for $\sin \alpha $ is
remarkably simple, namely 
\begin{equation}
\sin \alpha _{m}=\frac{\left| m_{+}-m_{-}\right| }{m+1}\mathrm{\
(exact~result~for~constant~\phi }_{m}\mathrm{).}  \label{exact}
\end{equation}
Numerically this is 0.74. For large $m$ the values of Eqs. (\ref{sinform})
and (\ref{exact}) are identical since the 1 in Eq.\ (\ref{exact}) is then
negligible.

The above results of $\alpha _{m}\rightarrow $ constant, and $\Phi _{m}=0,$
seem to imply that the ``hidden polarization direction,'' whose orientation
we are trying to determine, is at zero azimuthal angle and either above or
below the transverse plane at \emph{polar} angle $\alpha _{m}$. This angle
in turn is related to the angles $\pm \Phi _{0}$ of the peaks of $g_{m}(\Phi
)$ through the cone of equal probability (CEP) of Fig.~\ref{FirstCone} The
cone opening angle is $\Phi _{0}$ and the polar angle of the cone is the
complement $\alpha =\pi /2-\Phi _{0}$. A general position on the CEP around
the measurement angle $\phi _{1}=0$ obeys 
\begin{eqnarray}
P(\pm ) &=&\left| _{\pm }\left\langle 0|\beta ,\gamma \right\rangle \right|
^{2}=\frac{1}{2}\left[ 1\pm \sin \beta \cos \gamma \right] =\frac{1}{2}%
\left[ 1\pm \sin \alpha \right]  \nonumber \\
&=&\frac{1}{2}\left[ 1\pm \sin \left( \frac{\pi }{2}-\Phi _{0}\right)
\right] =\frac{1}{2}\left[ 1\pm \cos \Phi _{0}\right]
\end{eqnarray}
Thus all we can say about the hidden-spin angle so far is that $\sin \beta
\cos \gamma =\sin \alpha =\cos \Phi _{0}.$ We could have ($\beta ,\gamma
)=(\alpha $,$0)$ $,$ or ($\beta ,\gamma )=(\pi /2$,$\Phi _{0})$ or somewhere
else on the cone, but we cannot tell yet which it is. The distribution $%
g(\Phi )$ peaks at the two places where the CEP crosses the transverse plane.

We can analytically show the double peaking property of $g(\Phi )$ in the
case where all measurement angles $\phi _{i}$ are equal, say, to zero. Ref.~%
\onlinecite{CD} shows how this function has two equal sharp maxima in the
range we consider. For completeness we repeat this analysis in Appendix D,
getting a Gaussian approximation about each maximum. For large numbers of
measurements these Gaussians act as delta-functions at the two angles $\pm
\Phi _{0}$ centered on $\phi =0,$ as we have found in the experiment. Since
there are two delta-functions then we find from Eq.~(\ref{SpinIntegral})
that 
\begin{equation}
P_{m}(+)=\frac{1}{2}\left[ (1+\cos (+\Phi _{0}))+(1+\cos (-\Phi
_{0}))\right] =\frac{1}{2}\left( 1+\cos (\Phi _{0})\right) .
\end{equation}
From Eq.~(\ref{AngleSol}) in Appendix D, this is 
\begin{equation}
P_{m}(+)=\cos ^{2}\left( \Phi _{0}/2\right) =\frac{m_{+}}{m}.
\end{equation}
Of course, this is the same final value we got for $P_{m}(+)$ above in Eq.~(%
\ref{sinform}). We also comfirm the relation between $\alpha $ and $\Phi
_{0} $: 
\begin{eqnarray}
\cos \Phi _{0} &=&\sin \alpha  \nonumber \\
\Phi _{0} &=&\pi /2-\alpha  \label{eq21}
\end{eqnarray}
as we found above by considering the CEP. As far as the general hidden
direction ($\beta ,\gamma )$ is concerned, all we can say is $\sin \beta
\cos \gamma =\cos \Phi _{0}$ with some ambiguity as to signs or $\pi /2$'s
in the angles.

Let us see how this all works out in our experiment. In our first experiment
we have $\cos ^{2}\left( \Phi _{0}/2\right) =262/300=0.873$ so that $\Phi
_{0}=0.73.$ Thus the peaks of $g$ will be at $\pm 0.73\mathrm{\ rad}$ as we
have found experimentally and is shown in Fig.~\ref{ggSpin}.

Can we can determine \emph{both} the unknowns $\beta $ and $\gamma $ from
measurements and so remove the ambiguity between the two sides $\pm \Phi
_{0} $ of the cone? The only possible way to do this is to use other
measurement angles. If we continue to measure only in the transverse plane,
however, it seems likely we will determine only a combination of $\beta $
and $\gamma $ and not both$;$ we would need to go out of the transverse
plane to get both$. $ We now consider changes in measurement angle $\phi
_{1} $.

\subsection{ Changing the measurement angle:}

Suppose we now pick a $\phi _{m+1}\cong +\Phi _{0},$ (we continue to take
the original constant $\phi _{1}$ to be zero), that is, we measure as close
to the angle of one of the peaks as we can in the next measurement. We might
miss it by $\delta $. What this does is to eliminate completely (or almost)
one of the peaks. We show this in Fig.~\ref{ggStart}. When working at a
single angle we did not know on which side of the measurement direction the
emerging polarization was. By moving in that direction (or away from it), we
do determine the side. We can see this analytically by using the Gaussian
approximation developed in Appendix D for the first $m$ interations. If we
move to $+\Phi _{0}$ then the value of $g_{m+1}(\Phi )$ is 
\begin{eqnarray}
g_{m+1}(\Phi ) &\sim &\left[ e^{-\frac{1}{2}m(\Phi -\Phi _{0})^{2}}+e^{-%
\frac{1}{2}m(\Phi +\Phi _{0})^{2}}\right] (1+\eta _{m+1}\cos (\Phi -\Phi
_{0}+\delta ))  \nonumber \\
&\approx &\left( 1+\eta _{m+1}\cos \delta \right) e^{-\frac{1}{2}m(\Phi
-\Phi _{0})^{2}}+\left( 1+\eta _{m+1}\cos (2\Phi _{0}-\delta )\right) e^{-%
\frac{1}{2}m(\Phi +\Phi _{0})^{2}}
\end{eqnarray}
where $\delta $ is our measurement inaccuracy. Now if it turned out that $%
\eta _{m+1}=-1$ we will have most likely made the wrong choice (because if
we are close to the correct angle, it is most likely that we will get an $+1$
result) and the wrong peak will be almost completely \emph{eliminated}. If $%
\eta _{m+1}=+1,$ then the correct peak will be emphasized by 2 and the wrong
peak given a smaller coefficient. As Fig.~\ref{ggStart} shows, it works with
just a few measurements (15 in this case) at the new angle $-0.73$. We now
know on which side of 0 the azimuthal angle lies.

\subsection{Iterations and Convergence:}

Next consider what continuing to measure $m^{\prime }$ more times at this
new angle produces. If we have done a good guess at what the polarization
angle is then we will get mostly up spin results and as factor containing
two peaks from these last $m^{\prime }$ measurements (with very small
separation $2\delta )$ near $\Phi _{0}.$ Then we will have 
\begin{eqnarray}
g_{m+m^{\prime }}(\Phi ) &\sim &\left[ e^{-\frac{1}{2}m(\Phi -\Phi
_{0})^{2}}+e^{-\frac{1}{2}m(\Phi +\Phi _{0})^{2}}\right] \left[ e^{-\frac{1}{%
2}m^{\prime }(\Phi -\Phi _{0}+\delta )^{2}}+e^{-\frac{1}{2}m^{\prime }(\Phi
-\Phi _{0}-\delta )^{2}}\right]   \nonumber \\
&\sim &2e^{-\frac{1}{2}(m+m^{\prime })(\Phi -\Phi _{0})^{2}}.
\end{eqnarray}
We get a sharpening of the peak. (If $\delta $ is too large we might still
have two very closely-spaced peaks near $\Phi _{0}$.)

Suppose we now have only the one peak after $m$ iterations and it is near $%
\Phi =+\Phi _{0}.$ That is, 
\begin{equation}
g_{m}(\Phi )\sim e^{-\frac{1}{2}m(\Phi -\Phi _{0})^{2}}  \label{eq24}
\end{equation}
This result implies that the mean deviation in the measurements should go as 
$\sqrt{1/m}.$ We Fourier transform this Gaussian to find 
\begin{equation}
a_{1}^{m}=2a_{0}^{m}e^{-\frac{1}{2m}}\cos \Phi _{0};~\text{~}%
b_{1}^{m}=2a_{0}^{m}e^{-\frac{1}{2m}}\sin \Phi _{0}  \label{fouriergauss}
\end{equation}
so that 
\begin{equation}
\tan \Phi _{m}=\tan \Phi _{0};~\text{~}~A_{m}=e^{-\frac{1}{2m}}.
\label{eq26}
\end{equation}
and if we continue to make other measurements near this same angle $\Phi
=\Phi _{0}+\delta _{m},$ where $\delta _{m}$ is the error in our $m\mathrm{th%
}$ guess about the exact position of $\Phi _{0}$, then 
\begin{eqnarray}
P(+) &=&\frac{1}{2}\left[ 1+e^{-\frac{1}{2m}}\left( \cos (\Phi _{0}+\delta
_{m})\cos \Phi _{0}+\sin (\Phi _{0}+\delta _{m})\sin \Phi _{0}\right) \right]
\nonumber \\
&\approx &\frac{1}{2}\left[ 1+e^{-\frac{1}{2m}}\left( 1+\delta _{m}\sin \Phi
_{0}\cos \Phi _{0}\right) \right] \approx \frac{1}{2}\left[ 1+e^{-\frac{1}{2m%
}}\right] \rightarrow 1.
\end{eqnarray}
The equivalent \emph{simpleminded} point of view is that the spin that was
originally at polar angle $\alpha =\Phi _{0}$ and zero azimuthal angle is 
\emph{now} is at polar angle $\alpha \approx \pi /2$ and azimuthal angle $%
+\Phi _{0}.$ But this is still the equivalent for purposes of measurement to
a (half) cone of possible ``real'' 3D positions.

Eqs.~(\ref{eq24}) and (\ref{eq26}) make explicit predictions of how the
width of the peak and the amplitude will proceed with measurement number. In
order to test those approximate forms let us set up the double peak in 10
measurements, produce a single peak by measuring near one peak angle for
just 20 steps and then go back to the original measurement direction $\phi
_{m}=0$ for a few hundred measurements. In starting anew we develop a
different random phase, of course. The results are given in Figs.~\ref
{AmpBkToZero}-\ref{devBkToZero}. The approximate equations work very well.

For these last measurements, the Fourier transform of Eq.~(\ref{fouriergauss}%
) is valid and using it in Eq.~(\ref{a1b1Eq}) with $\phi _{m}=0$ we find 
\begin{equation}
P_{m}(+)=\frac{1}{2}\left[ 1+\frac{a_{1}^{m}}{2a_{0}^{m}}\right] =\frac{1}{2}%
\left[ 1+e^{-\frac{1}{2m}}\left( \cos \Phi _{0}\right) \right] .
\end{equation}
Comparing with the more general point of view at unknown 3D emerging phase
angle ($\beta ,\gamma ),$ given by Eq.~(\ref{genProb}) we get 
\begin{equation}
\sin \beta \cos \gamma =e^{-\frac{1}{2m}}\cos \Phi _{0}
\end{equation}
We might have $\beta =\Phi _{0}$ and $\gamma =0$ or $\beta =\pi /2$ and $%
\gamma =\Phi _{0}$ but we can never ``know'' absolutely. (\emph{Indeed in
conventional quantum mechanics the wave function is a mixture of all
possible phase angles that give the same probability }$P(+).)$ Staying in
the transverse plane never yields both the space angles, just this
combination. We have now found the maximal information on the emerging phase
angle that the present set of measurements can offer. Appendix A discusses
how to extend the analysis to angles out of the transverse plane. However,
we do not extend the simulation to that case here.

\section{Conclusions}

We have presented a simulated experiment where two Bose condensates, one
with spin up and the other with spin down, have been allowed to interfere.
The standard quantum mechanical view is that a Fock state $\left|
N_{+},N_{-}\right\rangle $ is a linear combination of phase states. As we
make a series of measurements of spin orientation we develop a state that is
a mixture of such Fock states having varying numbers of up and down spins,
but with constant total number of spins. As we proceed the state becomes a
narrower mixture of phase states and the experiment gradually changes the
state into nearly just one particular phase state. In convertional quantum
mechanics, the measurement process is considered to have created the final
phase value by continual wave function collapse. However, as one does the
experiment, one may have an equally appealing additional-variable view that
the phase emerging was built-in all along and the experiment simply revealed
what it was. For the experimentalist, one view works just as well as the
other. The formalism developed in Ref.~\onlinecite{Phase} and generalized in
Appendix A is particularly appropriate for showing this equivalence since it
expresses the results of measurements on a Fock up-down state through an
integral over the ``conjugate''\ variable, the relative phase $\Phi ~$of the
two states.\ The probability of any sequence of results then appears as a
sum over the phase, exactly as in theories with additional variables.\ Each
measurement produces a new initial state, with a new $\Phi $ distribution;
so that \ the change of this probability distribution can be obtained very
easily for any sequence of measurements as we have done in our simulation.\
Thus we have been able to observe the emergence of the additional variable
under the effect of successive quantum measurements. The Amherst experiment
is found to be an ideal example of an actual experiment along these lines.

LKB is UMR 8552 de l'ENS, du CNRS and de l'UniversitŽ Pierre et Marie Curie.

\section{Appendices}

\subsection{Extension to spin measurement at arbitrary directions}

In this section we generalize the calculations of Sec. II and Ref.~%
\onlinecite{Phase}, where the directions of spin measurements were assumed
to be in the transverse plane; we now assume that the directions are
arbitrary.\ Instead of a single angle $\phi _{i}$ to characterize each
measurement, we then need two angles $\theta _{i}$ and $\phi _{i}.$ (It is
more convenient to come back to the standard notation and keep $\theta _{i}$
for the polar angle; $\phi _{i}$ therefore now replaces the notation $\theta
_{i}$ of Ref.~\onlinecite
{Phase}.)\ We also re-introduce orbital variables, which were ignored in
Sec. II. We call $\left| \alpha \right\rangle $ and $\left| \beta
\right\rangle $ the two spins states (up and down) and $\Psi _{\alpha ,\beta
}(\mathbf{r)}$ the corresponding field operators.\ The spin component along
the direction of measurement is: 
\begin{equation}
\sigma _{\theta ,\phi }(\mathbf{r})=\cos \mathbf{\theta ~}\left[ \Psi
_{\alpha }^{\dagger }(\mathbf{r)}\Psi _{\alpha }(\mathbf{r)-}\Psi _{\beta
}^{\dagger }(\mathbf{r)}\Psi _{\beta }(\mathbf{r)}\right] +\sin \theta
~\left[ e^{-i\phi }~\Psi _{\alpha }^{\dagger }(\mathbf{r)}\Psi _{\beta }(%
\mathbf{r})+e^{i\phi }~\Psi _{\beta }^{\dagger }(\mathbf{r)}\Psi _{\alpha }(%
\mathbf{r)}\right]  \label{gene-1}
\end{equation}
and the projector\footnote{%
Strictly speaking, the projectors contain an integral over a small domain $%
\Delta_{\mathbf{r}}$, which we do not write explicitly here for the sake of
simplicity It is assumed that the domains are sufficiently small to make the
average number of particles in them much less than 1. See the discussion in
\S \ 1.1 of Ref.~\onlinecite{Phase}, after equation (5).} over the spin
eigenstates: 
\begin{equation}
P_{\eta }=\frac{1}{2}\left[ n(\mathbf{r})\mathbf{~+~}\eta \mathbf{~}\sigma 
\mathbf{_{\theta ,\phi }(\mathbf{r)}~}\right] ,  \label{gene-2}
\end{equation}
with $\eta =\pm 1$ and the local density is defined by: 
\begin{equation}
n(\mathbf{r)=~}\Psi _{\alpha }^{\dagger }(\mathbf{r)}\Psi _{\alpha }(\mathbf{%
r)+~}\Psi _{\beta }^{\dagger }(\mathbf{r)}\Psi _{\beta }(\mathbf{r).}
\label{gene-3}
\end{equation}
The projector is then proportional to: 
\begin{eqnarray}
P_{\eta }(\mathbf{r;}\theta ,\phi )~ &\sim &\left[ 1+\eta \cos \theta
\right] ~\Psi _{\alpha }^{\dagger }(\mathbf{r)}\Psi _{\alpha }(\mathbf{r)+~}%
\left[ 1-\eta \cos \theta \right] ~\Psi _{\beta }^{\dagger }(\mathbf{r)}\Psi
_{\beta }(\mathbf{r)}  \nonumber \\
&&+\eta \sin \theta ~\left[ e^{-i\phi }~\Psi _{\alpha }^{\dagger }(\mathbf{r)%
}\Psi _{\beta }(\mathbf{r)+}e^{i\phi }~\Psi _{\beta }^{\dagger }(\mathbf{r)}%
\Psi _{\alpha }(\mathbf{r)}\right] .  \label{gene-4}
\end{eqnarray}

As in \S 3.1 of Ref.~\onlinecite{Phase}, we assume that the system of spin
particles is initially in the state: 
\begin{equation}
\left| \Psi _{0}\right\rangle ~=~\left| N_{a}:u_{a},\alpha
;~N_{b}:u_{b},\beta \right\rangle  \label{gene-5}
\end{equation}
where the orbital states correspond to the wave functions (normalized to
one): 
\begin{equation}
\left\langle \mathbf{r}\right. \left| u_{a,b}\right\rangle =u_{a,b}(\mathbf{r%
}).  \label{gene-6}
\end{equation}
The initial system is therefore simply the juxtaposition of a large number $%
N_{a}$ of particles condensed into the wave function $u_{a}(\mathbf{r})$
with a large number $N_{b}$ of particles condensed into the wave function $%
u_{b}(\mathbf{r})$. We now assume that $M$ spin measurements are performed
at points $\mathbf{r}_{1}$, $\mathbf{r}_{2},\cdots ,\mathbf{r}_{m}$ in spin
direction defined by angles $(\theta _{1},\phi _{1})$, $(\theta _{2},\phi
_{2}),\cdots ,(\theta _{m},\phi _{m})$ and calculate the probability for
obtaining a series of results $\eta _{1}$, $\eta _{2},\cdots ,\eta _{m}$
(all $\eta $'s are equal to $\pm 1$).\ The corresponding probabity is the
average value in state $\left| \Psi _{0}\right\rangle $ of a product of $P$
projectors $P_{\eta _{i}}(\theta _{i},\phi _{i})$.

The rest of the calculation is very similar to that Ref.~\onlinecite{Phase};
the only difference being the presence of the terms in $\eta \cos \theta $
in Eq.~(\ref{gene-4}) and the $\sin \theta $ factor, which do not change
much the calculation.\ The same considerations apply on the conservation of
the number of particles in each state and, in the limit where $m\ll
N_{a},N_{b}$, we can express this conservation through an integral over a
phase $\Phi $. The probability then becomes proportional to the expression: 
\begin{eqnarray}
&&\int_{0}^{2\pi }\frac{d\Phi }{2\pi }~\prod_{i=1}^{m}\biggl\{ \left[ 1+\eta
_{i}\cos \theta _{i}\right] ~N_{a}\left| u_{a}(\mathbf{r}_{i})\right|
^{2}+\left[ 1-\eta _{i}\cos \theta _{i}\right] ~N_{b}\left| u_{b}(\mathbf{r}%
_{i})\right| ^{2}\biggr.  \nonumber \\
&&\biggl. +~\eta _{i}\sin \theta _{i}\sqrt{N_{a}N_{b}}\left[ u_{a}(\mathbf{r}%
_{i})u_{b}^{*}(\mathbf{r}_{i})e^{i\left( \phi _{i}-\Phi \right) }+\text{c.c}%
.\right] \biggr\} ,  \label{gene-7}
\end{eqnarray}
where c.c. stands for ``complex conjugate.''\ The second line ot this result
can also be written as 
\begin{equation}
+~2\eta _{i}\sin \theta _{i}~\sqrt{N_{a}N_{b}}\left| u_{a}(\mathbf{r}%
_{i})\right| \left| u_{b}(\mathbf{r}_{i})\right| ~\cos \left[ \xi (\mathbf{r}%
_{i})-\phi _{i}-\Phi \right]  \label{gene-8}
\end{equation}
where $\xi (\mathbf{r})$ is the relative phase of the two wave functions of
the condensates: 
\begin{equation}
\xi (r)=\arg \left[ u_{a}(\mathbf{r})/u_{b}(\mathbf{r})\right] .
\label{gene-9}
\end{equation}
This second line contains all the $\Phi $ dependence of the probabilities;
in other words it is the only origin of correlations between different
measurements. It also contains the $\mathbf{r}$ dependence of the
probability, which produces the fringes in space.

The discussion of \S \S\ 1.1 and 3.1 of Ref.~\onlinecite{Phase} shows that
effect of the $i$-th measurement on the $\Phi $ distribution is merely to
multiply this distribution by a function $D_{i}\left( \Phi \right) $ that is
nothing but the content of the $i$-th curly brackect in Eq. (\ref{gene-7})
(with a normalization constant that is unimportant for our discussion here).
This multiplication provides the evolution of the information on the
relative phase $\Phi $ that is obtained by this measurement. Whatever
measurement parameters $\mathbf{r}_{i}$, $(\theta _{i},\phi _{i})$ are
arbitrarily chosen, the multiplying function $D_{i}\left( \Phi \right) $ is
always the sum of a constant (first line of (\ref{gene-7})) plus a
sinusoidal variation given by (\ref{gene-8}).\ The former depends on $%
\mathbf{r}_{i}$ and $\theta _{i}$ only, with the $\mathbf{r}_{i}$ dependence
involving only the densities of probabilities associated with the condensed
states; the latter depends also on $\phi _{i}$ as well as the relative phase
of the two states. Both depend in general on the result of the measurement $%
\eta _{i}$, as opposed to the situation for measurement in transverse
directions only ($\theta _{i}=\pi /2$).

If, for instance, we assume that $\eta _{i}=+1$, the maximum and the minimum
of the contribution to the probability of the $i$-th measurement can be
written as 
\begin{equation}
\left\{ \sqrt{N_{a}}\left| u_{a}(\mathbf{r}_{i})\right| \cos \frac{\theta
_{i}}{2}\pm \sqrt{N_{b}}\left| u_{b}(\mathbf{r}_{i})\right| \sin \frac{%
\theta _{i}}{2}\right\} ^{2}.  \label{gene-10}
\end{equation}
The best contrast will therefore be obtained if the minimum vanishes, that
is if: 
\begin{equation}
\tan \frac{\theta _{i}}{2}=\sqrt{\frac{N_{a}}{N_{b}}~}\left| \frac{u_{a}(%
\mathbf{r}_{i})}{u_{b}(\mathbf{r}_{i})}\right| .  \label{gene-11}
\end{equation}
This provides the optimum value of $\theta _{i}$ for each measurement
position $\mathbf{r}$ if $\eta _{i}=+1$. But, if $\eta _{i}=-1$, it is easy
to see that $\cos \theta _{i}/2$ and $\sin \theta _{i}/2$ are interchanged
in (\ref{gene-10}), so that the the right hand side of (\ref{gene-11}) now
provides the inverse of $\tan \theta _{i}/2$; the optimum value of $\theta
_{i}$ corresponding to the two possible results are therefore symmetrical
with respect to the horizontal plane.

As a consequence, the choice of the optimum value of $\theta _{i}$ is not
easy in general, since it depends on the random result of an experiment that
is not yet known when the apparatus is adjusted.\ If, nevertheless, one can
locate the position $\mathbf{r}_{i}$ of the measurement at a point where the
two bosonic fields have the same intensity: 
\begin{equation}
N_{a}\left| u_{a}(\mathbf{r}_{i})\right| ^{2}~=~N_{b}\left| u_{b}(\mathbf{r}%
_{i})\right| ^{2}  \label{gene-12}
\end{equation}
the dilemma disappears: the optimum value of $\theta _{i}$ is $\pi /2$,
corresponding to a measurement performed in a direction of the transverse
plane.\ In this case, the flexibility introduced by the new parameter $%
\theta _{i}$ is useless.\ Nevertheless, condition Eq. (\ref{gene-12}) is not
necessarily easy to meet, and may even be impossible for some
configurations.\ Then, the optimization of the polar direction of
measurement becomes relevant, but only if one already has a good idea in
advance of what the most likely value of $\Phi $ is, so that with an
appropriate choice of $\varphi _{i}$ it is possible to infer what the result 
$\eta _{i}$ of the next measurement will be with a good probability.\ The
conclusion is then that the flexibility introduced by $\theta _{i}$ may be
useful, but only after a series a measurements has already been performed,
so that $\Phi $ is already reasonably well known.

\subsection{Proof that $\Phi _{m}=\phi $ (mod $\pi )$ for measurement along
a single direction}

Consider the situation in which the measurement angle $\phi $ remains the
same in every measurement. Then we can Fourier transform 
\begin{equation}
g_{m}(\Phi )=\prod_{i=1}^{m-1}(1+\eta _{i}\cos (\Phi -\phi )).
\end{equation}
in a series in $\cos (\phi -\Phi )$ rather than in $\cos \Phi $ and $\sin
\Phi .$ That is 
\begin{equation}
g_{m}(\Phi )=a_{0}^{m}+\sum_{q}c_{q}^{m}\cos (q(\Phi -\phi
))=a_{0}^{m}+\sum_{q}\left( c_{q}^{m}\cos \Phi \cos \phi +c_{q}^{m}\sin \Phi
\sin \phi \right) \mathrm{.}
\end{equation}
from which we find 
\begin{equation}
P_{n}(\mathbf{+})\sim 1+\frac{c_{1}^{m}}{2a_{0}^{m}}.
\end{equation}
By comparison with Eq.~(\ref{origg}) we see that 
\begin{equation}
a_{1}^{m}=c_{1}^{m}\cos \phi ;~\text{~}~b_{1}^{m}=c_{1}^{m}\sin \phi ,
\end{equation}
which, by Eq.~(\ref{eqn3}), shows that $\tan \Phi _{m}=\tan \phi $ or 
\begin{equation}
\Phi _{m}=\phi \mathrm{\ or~}\phi +\pi ,
\end{equation}
and that $c_{1}^{m}/2a_{0}^{m}=\sin \alpha _{m}\cos (\Phi _{m}-\phi )\mathrm{%
\ }=\pm \sin \alpha _{m}.$ This also gives 
\begin{equation}
P_{m}(\mathbf{+})=\frac{m_{+}}{m}=\frac{1}{2}(1\pm \sin \alpha _{m}).
\end{equation}
where the $+$ sign occurs with $m_{+}>$ $m_{-}$ and the $-$ sign in the
opposite case so that $\sin \alpha _{m}$ is positive.

\subsection{Analytic results for constant measurement angle}

When all measurement angles are the same value we can with full generality
take that angle equal to zero and write 
\begin{eqnarray}
g_{m}(\Phi ) &=&\prod_{i=1}^{m-1}(1+\eta _{i}\cos (\Phi ))=(1+\cos (\Phi
))^{m_{+}}(1-\cos (\Phi ))^{m-}  \nonumber \\
&=&2^{m}\left( \cos \frac{\Phi }{2}\right) ^{m_{+}}\left( \sin \frac{\Phi }{2%
}\right) ^{m_{-}}
\end{eqnarray}
Then the Fourier transforms are found from integral tables\cite{Abram} to be 
\begin{eqnarray}
a_{0}^{m} &=&\frac{2^{m}}{2\pi }\int_{0}^{2\pi }d\Phi \left( \cos \frac{\Phi 
}{2}\right) ^{m_{+}}\left( \sin \frac{\Phi }{2}\right) ^{m_{-}}  \nonumber \\
&=&\frac{2^{m}}{\pi }\frac{\Gamma \left( m_{+}+\frac{1}{2}\right) \Gamma
\left( m_{-}+\frac{1}{2}\right) }{\Gamma \left( m+1\right) }
\end{eqnarray}
\begin{eqnarray}
a_{1}^{m} &=&\frac{2^{m}}{\pi }\int_{0}^{2\pi }d\Phi \left( \cos \frac{\Phi 
}{2}\right) ^{m_{+}}\left( \sin \frac{\Phi }{2}\right) ^{m_{-}}\cos \Phi 
\nonumber \\
&=&\frac{2^{m+1}}{\pi }\frac{\Gamma \left( m_{+}+\frac{3}{2}\right) \Gamma
\left( m_{-}+\frac{1}{2}\right) -\Gamma \left( m_{+}+\frac{1}{2}\right)
\Gamma \left( m_{-}+\frac{3}{2}\right) }{\Gamma \left( m+2\right) }
\end{eqnarray}
with $b_{1}^{m}=0$ by symmetry. Taking the ration of these and using the
definition of $\sin \alpha _{m}$ gives 
\begin{equation}
\sin \alpha _{m}=\frac{\left| a_{1}^{m}\right| }{2a_{0}^{m}}=\frac{\left|
m_{+}-m_{-}\right| }{m+1}.
\end{equation}

This result allows us to understand the behavior of $A_{m}$ as seen in Fig.
1. We notice that the plot has small upward curves followed by abrupt
downward jumps. We find that the upward sweep is a sequence of all $\eta =1$
results while a downward jump is a single $\eta =-1.$ Suppose, as occurs in
the figure that $m_{+}$ is considerably larger than $m_{-}.$ Then when a new
up spin result occurs the change in $A_{m}$ is given by 
\begin{equation}
\Delta _{+}=\frac{m_{+}+1-m_{-}}{m+2}-\frac{m_{+}-m_{-}}{m+1}=\frac{2m_{-}+1%
}{\left( m+2\right) \left( m+1\right) }
\end{equation}
while, if a down spin occurs in the measurement, the change in this quantity
is 
\begin{equation}
\Delta _{-}=\frac{m_{+}-m_{-}-1}{m+2}-\frac{m_{+}-m_{-}}{m+1}=-\frac{2m_{+}+1%
}{\left( m+2\right) \left( m+1\right) }
\end{equation}
The second quantity, the jump down, is much greater in magnitude than the
previous upward move, giving the peculiar shape of the curve.

\subsection{The peaks in $g(\Phi )$ for the spin phase}

Here we analyze $g(\Phi )$ in the case of all equal measurement axes $\phi
_{i}=\phi $ to show it has two sharp peaks. We can take the single
measurement angle to be zero with generality (or work with $\Phi ^{\prime
}=\Phi -\phi ).$ The location of these peaks corresponds to what we have
found numerically. Write $g(\Phi )$ in the half-angle form 
\begin{equation}
g(\Phi )=\left( \cos ^{2}\left[ (\Phi )/2\right] \right) ^{m_{+}}\left( \sin
^{2}\left[ (\Phi )/2\right] \right) ^{m_{-}}.
\end{equation}
Set the derivative of the logarithm of this to zero to determine the
position of the maxima. 
\begin{equation}
\frac{d\ln g(\Phi )}{d\Phi }=-m_{+}\frac{\sin \left[ (\Phi )/2\right] }{\cos
\left[ (\Phi )/2\right] }+m_{-}\frac{\cos \left[ (\Phi )/2\right] }{\sin
\left[ (\Phi )/2\right] }=0,
\end{equation}
which is equivalent to 
\begin{equation}
\cos ^{2}\left[ (\Phi )/2\right] =\frac{m_{+}}{m}.
\end{equation}
This equation has two solutions in the range $0\leq \Phi <2\pi .$ These are
at 
\begin{equation}
\Phi =\pm \Phi _{0}
\end{equation}
where $\Phi _{0}$ is in the range $[0,\pi ]$ or 
\begin{equation}
\cos ^{2}\left[ \Phi _{0}/2\right] =\frac{m_{+}}{m}.  \label{AngleSol}
\end{equation}
Thus the two peaks occur symmetrically about the value of $\Phi =0$ just as
we have found ``experimentally.'' Correspondingly we have 
\begin{equation}
\sin ^{2}\left[ \Phi _{0}/2\right] =\frac{m_{-}}{m}.
\end{equation}

We get a Gaussian approximation to $g$ by Taylor expanding $\ln g$ about $%
\pm \Phi _{0}.$ We use 
\begin{equation}
\left. \frac{d^{2}\ln g(\Phi )}{d\Phi ^{2}}\right| _{\Phi _{0}}=-m,
\end{equation}
so that\cite{CD} 
\begin{equation}
g(\Phi )=\left( \frac{m_{+}}{m}\right) ^{m_{+}}\left( \frac{m_{-}}{m}\right)
^{m_{-}}\left[ e^{-\frac{1}{2}m(\Phi -\Phi _{0})^{2}}+e^{-\frac{1}{2}m(\Phi
+\Phi _{0})^{2}}\right] .
\end{equation}
For large $m$ these act like delta-functions.

\newpage

\begin{figure}[h]
\centering
\includegraphics[width=4in, height=2.38in]{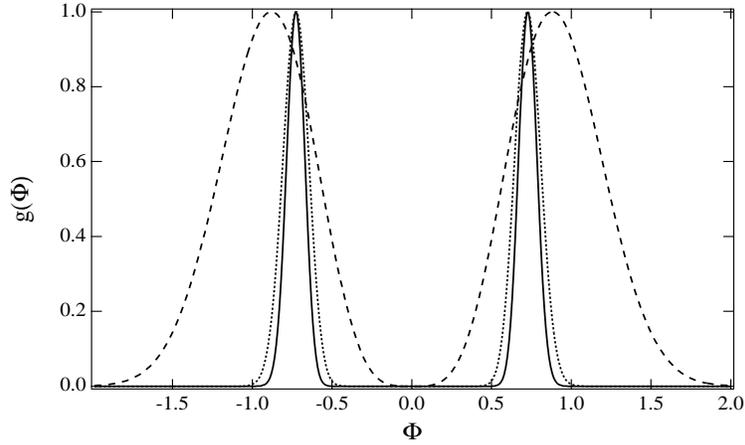}
\caption{The angular distribution $g(\Phi )$ as a function of angle for
three iteration step lengths, 10 steps (dashed line), 150 steps (dottted
line), and 300 steps (solid line). For a single measuring angle this always
has two equal peaks corresponding to the intersection of the spin cone with
the transverse plane. The peaks narrow with step length.}
\label{ggSpin}
\end{figure}

\begin{figure}[h]
\centering
\includegraphics[width=4in, height=2.52in]{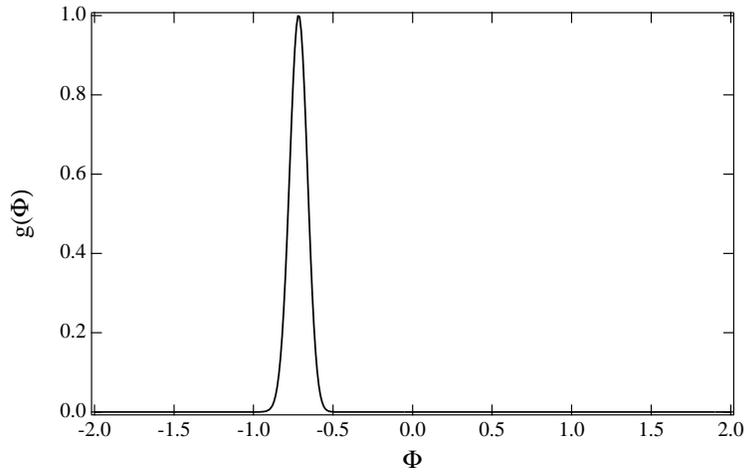}
\caption{Typical $g(\Phi )$ after the second peak has been eliminated by a
few measurements at the positive peak.}
\label{ggStart}
\end{figure}

\begin{figure}[h]
\centering
\includegraphics[width=4in, height=2.42in]{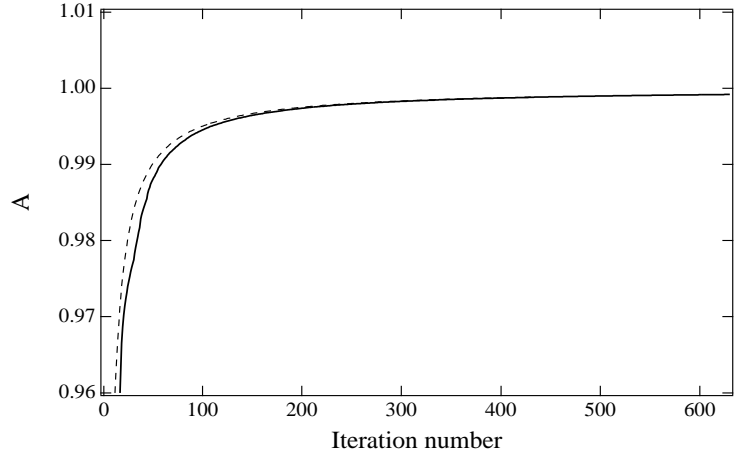}
\caption{The amplitude $A_{m}$ as a function of interation step when all
measurements, after eliminating the second peak of $g(\Phi )$ early on, are
back at the original measurement angle $\phi_{1} =0$. The dotted line is the
approximation $e^{-1/2m}$.}
\label{AmpBkToZero}
\end{figure}

\begin{figure}[h]
\centering
\includegraphics[width=4in, height=2.79in]{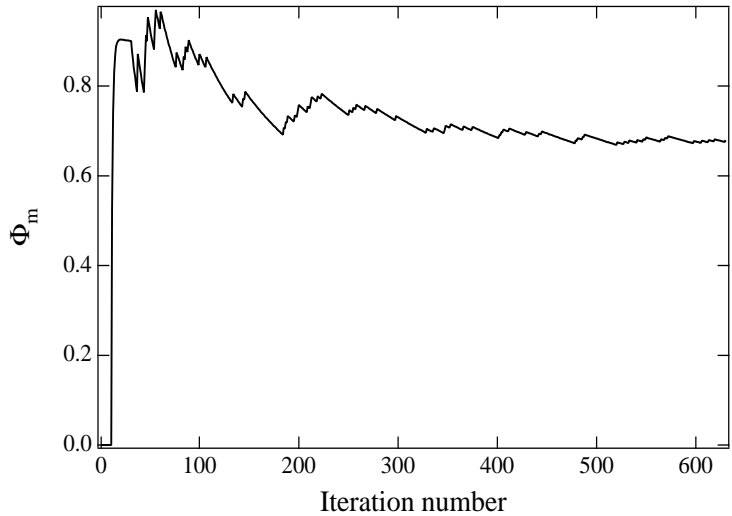}
\caption{The azimuthal angle $\Phi _{m}\approx \Phi_{0}$ as a function of
interation step when all measurements, after eliminating the second peak of $%
g(\Phi )$ early on, are back at the original measurement angle $\phi_{1} =0$.}
\label{phiBkToZero}
\end{figure}

\begin{figure}[h]
\centering
\par
\includegraphics[width=4in, height=2.52in]{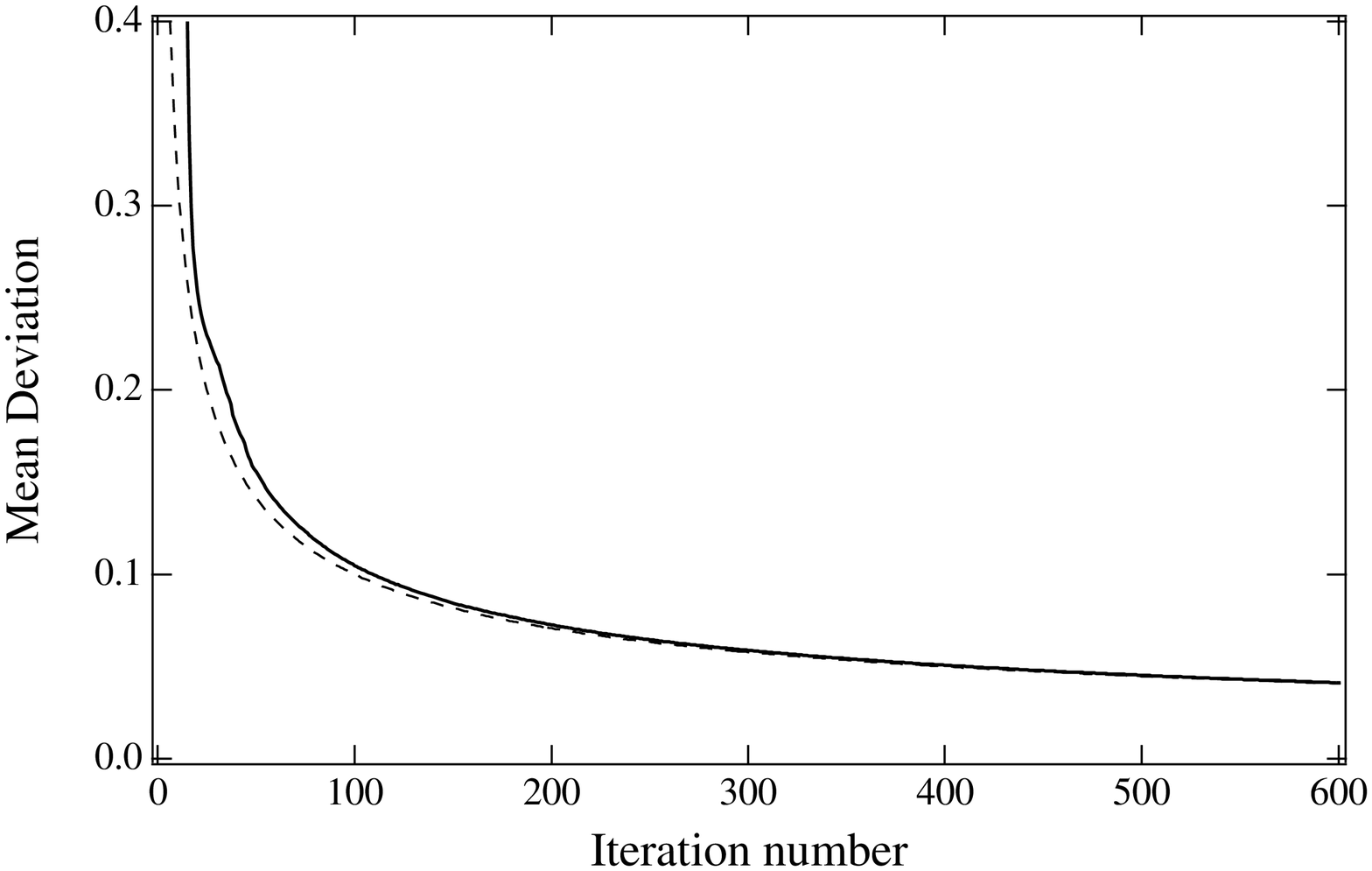} .
\caption{The mean deviation (width of $g(\Phi )$) as a function of
interation step when all measurements, after eliminating the second peak of $%
g(\Phi )$ early on, are back at the original measurement angle $\phi_{1} =0$.
The dotted line is the approximate form $1/\protect\sqrt{m}$}
\label{devBkToZero}
\end{figure}

\begin{thebibliography}{99}
\bibitem{JILA}  H.\ J.\ Lewandowski, D.\ M.\ Harber, D.\ L.\ Whitaker, and
E.\ A.\ Cornell, Phys. Rev. Lett. \textbf{88}, 070403 (2002).

\bibitem{JILA2}  J.\ M.\ McGuirk, H.\ L.\ Lewandowski, D.\ M.\ Harper, T.\
Nikuni, J.\ E.\ Williams, and E.\ A.\ Cornell, Phys. Rev. Lett. \textbf{89},
090402 (2002).

\bibitem{Jammer}  M.\ Jammer, ``The conceptual development of quantum
mechanics'', McGraw Hill, 2nd ed. (1989).

\bibitem{Hall}  Mark H.\ Wheeler, Kevin M.\ Mertes, Jessie D.\ Erwin, and
David S.\ Hall, Phys. Rev. Lett.\textbf{\ 93}, 170402-1 (2004).

\bibitem{De Broglie}  L.\ de Broglie, in ``Rapport au V\`{e}me congr\`{e}s
de physique Solvay'', Gauthier Villars, Paris (1930).

\bibitem{Bohm}  D.\ Bohm, Phys.\ Rev. \textbf{85}, 166 and 180 (1952).

\bibitem{EPR}  A.\ Einstein, N.\ Rosen and B.\ Podolsky, Phys.\ Rev.\ 
\textbf{47}, 777 (1935); or in ``Quantum theory of measurement'', J.A.\
Wheeler and W.H.\ Zurek eds, Princeton University Press (1983), 138.

\bibitem{Bell1}  J.\ S.\ Bell, ``On the Einstein-Podolsky-Rosen paradox'',
Phyics \textbf{1}, 195 (1964); reprinted in ``Quantum theory of
measurement'', J.A.\ Wheeler and W.H.\ Zurek eds, Princeton Univ.\ Press
(1983), 396.

\bibitem{Bell2}  J.\ S.\ Bell, ``Speakable and unspeakable in quantum
mechanics'', Cambridge University Press (1987).

\bibitem{footnote}  The so called ``efficiency loophole''\ has not yet been
closed, but very few physicists now think that quantum mechanics will fail
in an experiment that is free of this loophole.

\bibitem{Mermin}  N.\ D.\ Mermin, Rev.\ Mod.\ Phys. \textbf{65}, 803 (1993).

\bibitem{AJP}  F.\ Lalo\"{e}, Am.\ J.\ Phys.\ \textbf{69}, 655 (2001).

\bibitem{Goldstein}  S.\ Goldstein, ``Quantum Theory without observers'',
Physics Today \textbf{51}, 42 (March 1998) and 38 (April 1998).

\bibitem{Anderson}  P.\ W.\ Anderson, Rev.\ Mod.\ Phys. \textbf{38}, 298
(1966); P.W.\ Anderson, ``Basic notions in condensed matter physics'',
Benjamin-Cummins (1984).

\bibitem{Javanainen}  J.\ Javanainen and Sun Mi Yoo, Phys.\ Rev.\ Lett. 
\textbf{76}, 161 (1996).

\bibitem{WCW}  T.\ Wong, M.J.\ Collett and D.\ F.\ Walls, Phys.\ Rev.\ 
\textbf{A 54}, R3718 (1996).

\bibitem{CGNZ}  J.\ I.\ Cirac, C.\ W.\ Gardiner, M.\ Naraschewski and P.\
Zoller, Phys.\ Rev.\ \textbf{A 54}, R3714 (1996).

\bibitem{CD}  Y.\ Castin and J.\ Dalibard, Phys.\ Rev.\ \textbf{A 55}, 4330
(1997).

\bibitem{M1}  K.\ M.\ lmer, Phys.\ Rev. \textbf{A 55}, 3195 (1997).

\bibitem{M2}  K.\ M.\ lmer, J.\ Mod.\ Opt. 44, 1937 (1997).

\bibitem{Phase}  F.\ Lalo\"{e}, \ss\ Eur.\ Phys.\ J. \textbf{33}, 87 (2005).

\bibitem{Bell3}  J.\ S.\ Bell, ``Are there quantum jumps?''\ in Ref.\ %
\onlinecite{Bell2}.

\bibitem{Ketterle}  M. R. Andrews, C. G. Townsend, H.-J. Miesner, D. S.
Durfee, D. M. Kurn, and W. Ketterle, Science \textbf{275}, 637 (1997).

\bibitem{Wallis}  H. Wallis, A. R\"{o}hrl, M. Naraschewski, and A. Schenzle,
Phys. Rev. A \textbf{53}, 2109 (1997).

\bibitem{Abram}  M.\ Abramowitz and I.\ A.\ Stegun, \textit{Handbook of
Mathematical Functions, }Dover Publications (1972), Sec. 6.2.
\end{thebibliography}
\end{document}